\documentstyle[prl,aps,eqsecnum,epsfig,multicol]{revtex}

\def\simge{\stackrel{>}{\sim} }

\def\Journal#1#2#3#4{{#1}{\bf #2}, #3 (#4)}

\def\NIMA{{Nucl. Instrum. Methods}~{\bf A}}
\def\NPA{{Nucl. Phys.}~{\bf A}}
\def\NPB{{Nucl. Phys.}~{\bf B}}
\def\PLB{{Phys. Lett.}~{\bf B}}

\def\PRL{Phys. Rev. Lett.\ }
\def\PRD{{Phys. Rev.}~{\bf D}}
\def\PRC{{Phys. Rev.}~{\bf C}}
\def\ZPC{{Z. Phys.}~{\bf C}}

\begin{document}
\draft

\title{Centrality dependence of $\pi^{+/-}$, $K^{+/-}$, $p$ and
 $\overline{p}$ production from $\sqrt{s_{_{NN}}}=130$~GeV
       Au + Au collisions at RHIC}

\author{
K.~Adcox,$^{40}$
S.{\,}S.~Adler,$^{3}$
N.{\,}N.~Ajitanand,$^{27}$
Y.~Akiba,$^{14}$
J.~Alexander,$^{27}$
L.~Aphecetche,$^{34}$
Y.~Arai,$^{14}$
S.{\,}
% ...
\\(PHENIX Collaboration)\\
}
\address{
% ...
$^{3}$Brookhaven National Laboratory, Upton, NY 11973-5000, USA\\
% ...
$^{14}$KEK, High Energy Accelerator Research Organization, Tsukuba-shi, Ibaraki-ken 305-0801, Japan\\
% ...
$^{27}$Chemistry Department, State University of New York - Stony Brook, Stony Brook, NY 11794, USA\\
% ...
$^{34}$SUBATECH (Ecole des Mines de Nantes, IN2P3/CNRS, Universite de Nantes) BP 20722 - 44307, Nantes-cedex 3, France\\
% ...
$^{40}$Vanderbilt University, Nashville, TN 37235, USA\\
% ...
}

\date{\today}        
\maketitle
\begin{abstract}

Identified $\pi^{+/-}$, $K^{+/-}$, $p$ and
 $\overline{p}$ transverse momentum spectra at mid-rapidity in  
$\sqrt{s_{NN}}$~=~130~GeV Au-Au collisions were measured by the PHENIX 
experiment at RHIC as a function of collision centrality. 
Average transverse momenta increase with the number of participating 
nucleons in a similar way for all particle species. 
Within errors, all mid-rapidity particle yields per participant are found to 
be increasing with the number of participating nucleons. There is an 
indication that $K^{+/-}$, $p$ and $\overline{p}$ yields per participant 
increase faster than the $\pi^{+/-}$ yields. In central collisions at high 
transverse momenta ($p_T \simge 2$~GeV/c), $\overline{p}$ and $p$ yields are 
comparable to the $\pi^{+/-}$ yields.

\end{abstract}


\begin{multicols}{2}   
\narrowtext            

We report first results on identified $\pi^{+/-}$, $K^{+/-}$,
$p$ and $\overline{p}$ production as a function of collision
centrality in $\sqrt{s_{NN}}$~=~130~GeV Au+Au collisions, measured
by the PHENIX experiment at the Relativistic Heavy 
Ion Collider (RHIC). The PHENIX objective is to search for signatures 
of deconfinement and chiral symmetry restoration and to study the 
transition from normal to deconfined nuclear matter by utilizing a 
wide variety of probes. 

Early RHIC results show that the transverse energy density and  
particle multiplicities are considerably higher than previously 
observed in relativistic heavy-ion collisions~\cite{PHENIX1,PHENIX2}. 
Measured energy densities extend into the region predicted to be 
favorable for the formation of a quark-gluon plasma\cite{QMproc}. 
Identified hadron spectroscopy provides a tool for studying reaction 
dynamics beyond that of global event characterization. The yields of hadrons 
reflect the particle production mechanism, while spectral shapes are 
sensitive to the dynamical evolution of the system. The mass and centrality 
dependence of the spectra can help differentiate between competing 
theoretical descriptions such as collective hydrodynamical 
expansion~\cite{Teaney,Kolb} or transverse momentum ($p_{T}$) broadening 
in the partonic stage of the reaction~\cite{Jurgen}. Additionally the 
relative yields of baryons and mesons at high ($p_{T}\simeq~2$~GeV/c) may 
give insight into baryon number transport~\cite{Vitev} and the interplay 
between soft and hard processes. 

The PHENIX detector has diverse particle
identification (PID) capabilities~\cite{Ham}, including excellent hadron
identification over a broad momentum range. This measurement was performed 
using a portion of the east central-arm spectrometer, covering pseudo-rapidity 
$|\eta|<0.35$ and $\Delta\phi=\pi/4$ in azimuthal angle. 

The collision $z$-vertex and the timing system's start signal are 
generated by the Beam-Beam Counters (BBC); two arrays 
of quartz Cherenkov radiators which surround the beam axis covering 
$\eta=\pm(3.0-3.9)$.
The tracking system includes a multi-layer focusing drift chamber located
outside an axially-symmetric magnetic field  at a radial distance between 
2.0\,m and 2.4\,m followed by a multi-wire proportional chamber with 
pixel-pad readout (PC1)~\cite{TEXAS}. 
Pattern recognition in the DC is based on 
a combinatorial Hough transform in the track bend plane~\cite{NIM}. 
The polar angle of the track is determined by  PC1 and  the 
collision $z$-vertex. A track model based on a field-integral 
look-up table determines the charged particle momentum and the path length to 
the time-of-flight (TOF) wall. The momentum resolution is 
$\delta p/p \simeq 0.6\% \oplus 3.6\%\ p$~(GeV/c).

The timing system stop signal for each particle is measured by the 
TOF scintillator wall, located at a radial distance of 5.06 m,
resulting in a flight-time measurement with a resolution of  
$\sigma\simeq115$\,ps. Reconstructed tracks are projected to the TOF and 
matched with hits in the scintillator slats 
using a momentum-dependent search window determined by 
multiple scattering and the momentum resolution. 
A velocity dependent energy loss cut based on a Bethe-Bloch 
parameterization is applied to the measured TOF pulse height.
Combining the momentum and flight-time, we reconstruct the particle mass and 
select particles by applying $2\sigma$ momentum dependent cuts in mass-squared.

Corrections for geometrical acceptance, decay-in-flight, 
momentum resolution and reconstruction efficiency are determined using a 
single-particle full GEANT Monte Carlo (MC) simulation. The acceptance 
correction assumes that the spectra are flat in azimuth and in rapidity for 
$|y|<0.5$~\cite{PHOBOS2}. Fiducial area cuts, energy loss and hit-track 
matching cuts are applied consistently in simulation and data. In peripheral 
events the track reconstruction efficiency is $\approx98\%$. As the 
centrality increases, the efficiency is reduced due to the increased detector 
occupancy. Multiplicity dependent corrections are obtained by embedding 
simulated tracks into real events. Track-by-track corrections are applied 
taking into account the event centrality and the particle species. The 
efficiency for $\pi^{+/-}$ in the most central events is $\approx(68\pm6)\%$, 
independent of momentum. In the case of overlapping hits in the TOF 
wall, the earliest pulse reaching each photo-multiplier is recorded. This 
favors the faster particles; hence, in central events heavier particles 
suffer an additional reconstruction inefficiency of $\approx4\%$. Corrections 
for feed-down from weak decays are not applied. A MC simulation is used to 
estimate the probability for reconstructing protons from $\Lambda$ decays as 
prompt protons. Within the PHENIX acceptance this probability is $\approx50\%$ at $p_{T}=0.5$\,GeV/c, $\approx32\%$ at $p_{T}=1$\,GeV/c and $\approx12\%$ 
at $p_{T}\ge2$\,GeV/c. Taking  ${{\Lambda}\over{p}}=1$ as an upper limit, we 
estimate  $33\%$ as the upper limit of weak decay contribution to the 
reported  $p$, $\overline{p}$ yields and maximal $16\%$ change of 
measured $<\!\!p_{T}\!\!>$.   
  
About 140,000 minimum bias events, representing $92\pm 4\%$ of the total 
inelastic cross-section of $6.8$\,b~\cite{PHENIX1} were analyzed. 
This sample was subdivided into five centrality classes: $0-5\%$, 
$5-15\%$, $15-30\%$, $30-60\%$ and $60-92\%$, using the BBC and 
Zero-Degree-Calorimeters for event characterization~\cite{PHENIX1}.
For each class, the average number of nucleons participating in the 
collision ($N_{part}$) is obtained from a  Glauber model calculation~\cite{ppg003}.
   
\begin{figure}
\centerline{\epsfig{file=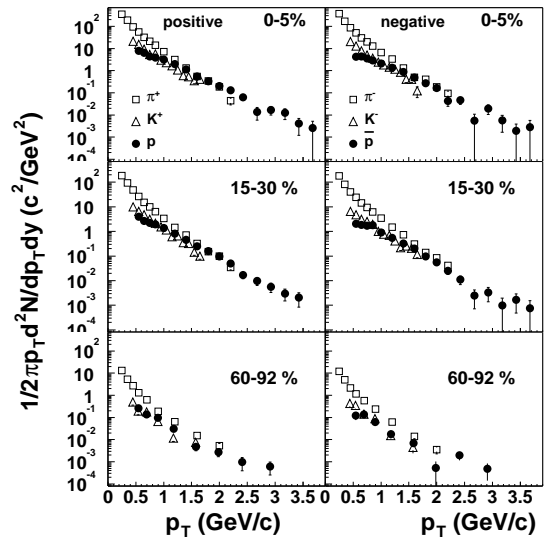,width=8.5cm}}
\caption[]{Transverse momentum spectra measured at mid-rapidity 
for $\pi^{+}, K^{+}, p$ (left) and $\pi^{-}, K^{-}, \overline{p}$  
at the three different centrality selections  indicated in each panel.
 The symbols indicated in the top panels apply for 
all centrality selections.
\label{f:fig1}}
\end{figure}

Fig.~\ref{f:fig1} shows the invariant yield as a function of $p_{T}$
 for $\pi^{+}, K^{+}, p$ ~(left panel) and   $\pi^{-}, K^{-},
\overline{p}$ for 
three centrality selections. Error bars in the figure are the 
combined statistical errors in the data and the corrections. Systematic 
uncertainties from acceptance, multiplicity-dependent 
efficiency corrections and PID cuts result in an overall 
systematic error in the absolute normalization of $\approx11\%$ for 
all species. As a consistency check, we have added all identified 
charged  hadron spectra in the $p_{T}$ region where PID is available for all 
species and compared to the PHENIX charged hadron measurement~\cite{ppg003}. 
The results agree to better than $10\%$ for all centralities. Additionally, 
the  PHENIX $\pi^{0}$ spectra~\cite{ppg003} and the $\pi^{+/-}$ spectra in 
the region of overlapping $p_{T}$ are within $\approx 10\%$ for the central 
and $\approx25\%$ for the peripheral selections, which is within the 
systematic errors of the two measurements. $\overline{p}$ results presented 
here are also in agreement (within errors) with recent publication~\cite{STAR2}.  
 
In peripheral  events the $\pi^{+/-}$ spectra 
exhibit a concave shape, well described by a power-law parameterization as 
observed in hadron-hadron collisions~\cite{UA1}. With increasing centrality 
the curvature in the spectra decreases, leading to an almost exponential 
dependence on $p_{T}$ for the most central events. 
Over the measured $p_{T}$ range, the $K^{+/-}$ spectra can be described by an 
exponential distribution either in $p_{T}$ or in 
$m_{T}=\sqrt{p_{T}^{2}+m^{2}}$, while the  $p$ and $\overline{p}$   
spectra can be described either as a Boltzmann or an exponential distribution 
in $m_{T}$. The slopes of the $m_{T}$ spectra flatten and the mean transverse 
momentum ($<\!\!p_{T}\!\!>$) increases with particle mass and with centrality.
This behavior has been previously observed in lower energy heavy ion 
collisions at the BNL-AGS~\cite{E866} and at the CERN-SPS~\cite{NA441,NA49} 
and was attributed to collective radial motion (flow). 

At lower energies, it is not uncommon for the proton yields to equal or 
exceed the $\pi^{+}$ yields, since many of the protons come from the initial 
state. A new feature observed for the first time at RHIC is that in central 
collisions at $p_{T}\approx2$\,GeV/c the  $\overline{p}$ yields are comparable 
to the $\pi^{-}$ yields. Positive and negative hadrons behave in a 
similar way. In central events the proton yields approach the $\pi^{+}$ 
spectra around $p_{T}\simeq 1.6$\,GeV/c.  As the centrality decreases, this 
happens at larger $p_{T}$. In peripheral events, $p$ and $\overline{p}$ 
spectra are below the $\pi^{+/-}$ spectra over the whole measured $p_{T}$ 
range. Since anti-protons are not as numerous as the protons,  
$\overline{p}$ and  $ \pi^{-}$ yields become comparable only at the high end 
of the  measured pion $p_{T}$ range in the most central collisions.  
We note that in $pp$~\cite{ISR} and 
$p\overline{p}$~\cite{FERMILAB} collisions at $\sqrt{s}=23-63$\,GeV
and $\sqrt{s}=300-1800$\,GeV respectively, the $\overline{p}/\pi^{-}$ ratio 
steadily increases with $p_{T}$ up to $\simeq 0.33$ at $p_{T}\simeq1.5$\,GeV/c 
nearly independent of $\sqrt{s}$. Data on baryon/meson ratios at higher 
$p_{T}$ are only available from $pp$ collisions at ISR energies 
($\sqrt{s}=23- 63$\,GeV) and show that above $p_{T}\simeq1.5$\,GeV/c the 
$\overline{p}/\pi^{-}$ ratio rises to $\simeq 0.4 $ at $p_{T}\simeq2$\,GeV/c 
and then drops, as expected if valence quark jet fragmentation is the 
dominant production mechanism for high $p_{T}$ hadrons. In central collisions 
at RHIC the high $p_{T}$ (anti)proton/pion ratios are $\sim~1$ -  
much larger than in $pp$ collisions.

Hydrodynamic calcualtions with fixed freeze-out temperature\cite{Kolb} or 
with freeze-out modeled using a hadronic cascade\cite{Teaney} suggest that 
hydrodynamic expansion is responsible for the baryon dominance at high 
$p_{T}$. However, protons/anti-protons produced via a baryon junction 
mechanism combined with jet-quenching in the pion channel are shown to 
exhibit the same effect~\cite{Vitev}. Intrinsic $p_{T}$ broadening in the 
partonic phase caused by gluon saturation\cite{Jurgen} gives yet another 
alternative explanation. The above models have similar predictions in the 
$p_{T}$ range measured here, but show different behavior at higher $p_{T}$. 
New data with broader $p_{T}$ range is needed in order to distinguish 
between currently available theories. 
    
\begin{figure}
\centerline{\epsfig{file=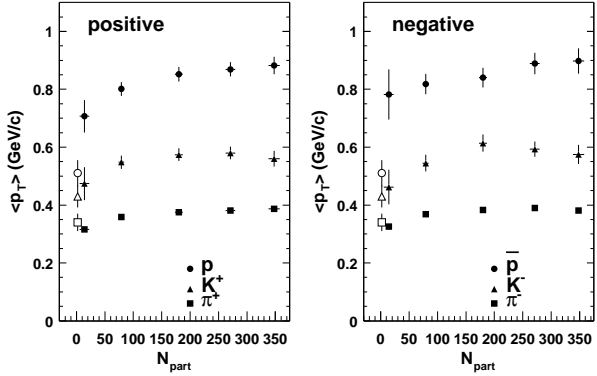,width=8.5cm}}
\caption[]{Average transverse momentum for $\pi^{+}, K^{+}, p$ (left) and 
$\pi^{-}, K^{-}, \overline{p}$ as a function of the number of nucleons 
participating in the collision $N_{part}$. The error bars represent the 
statistical errors. The systematic errors are discussed in the
text. The open points are interpolations from $pp$ and $p\overline{p}$
data, see text for details. 
\label{f:fig2}}
\end{figure}

\begin{figure}
\centerline{\epsfig{file=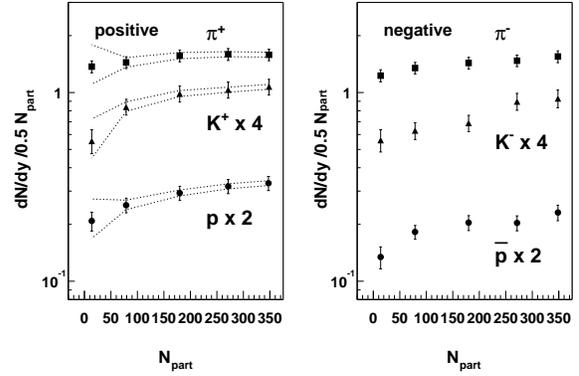,width=8.5cm}}
\caption[]{$dN/dy_{|y=0}$ per participant 
for $\pi^{+}, K^{+}, p$ (left) and $\pi^{-}, K^{-}, \overline{p}$ as 
a function of $N_{part}$. The error bars include statistical and 
systematic errors in $dN/dy$.  The dashed lines around the positive hadrons 
show the effect of the systematic error on $N_{part}$ which affects all 
curves in the same way.
\label{f:fig3}}
\end{figure}

To quantify the centrality and mass dependencies of hadron production,  we 
determine the mean ($<\!\!p_{T}\!\!>$ shown in
Fig.~\ref{f:fig2}) and the integral ($dN/dy$ shown in
Fig.~\ref{f:fig3}) of the $p_{T}$ distributions. Both quantities require 
extrapolation of the spectra below and above the measured range. For each 
particle species, we use at least two functional forms consistent with the 
data, as outlined above and the results presented are averaged between two 
fits. The fraction of the yield in the extrapolated regions is
estimated  $30\pm 6 \%$ for $\pi^{+/-}$, $40\pm 8\%$ for $K^{+/-}$ and 
$25 \pm 7.5\%$ for $p$/$\overline{p}$. Combining systematic uncertainties 
in the extrapolation fractions and the estimated background under
mass-squared peaks ($2\%$, $5\%$ and $3\%$) yield systematic uncertainties  
in  the measured $<\!\!p_{T}\!\!>$ of $7\%$, $10\%$ and $8\%$  
for $\pi^{+/-}$, $K^{+/-}$ and $p$/$\overline{p}$, respectively. Further 
combining the above uncertainties, which only affect the shape of the spectra,
 with the $11\%$ uncertainty in the absolute normalization gives systematic
uncertainties of $13\%$, $15\%$ and $14\%$ in the measured $dN/dy$ for 
$\pi^{+/-}$, $K^{+/-}$ and  $p$/$\overline{p}$, respectively. We note that 
after converting to  $dN/d\eta$, the sum of the identified charged hadron 
yields agrees with the previously published PHENIX results on total charged 
multiplicity $dN_{ch}/d\eta$~\cite{PHENIX1} within $5\%$.  

Fig.~\ref{f:fig2} shows the $<\!\!p_{T}\!\!>$ as a function of $N_{part}$ 
for $\pi^{+}, K^{+}, p$~(left panel) and 
$\pi^{-}, K^{-}, \overline{p}$. Filled points are this measurement; 
open points at $N_{part}=2$ are interpolations to $\sqrt{s}=130$\,GeV
obtained from $pp$ and $p\overline{p}$ data at lower~\cite{ISR} and higher 
energies~\cite{FERMILAB}, respectively. In peripheral Au-Au collisions at 
RHIC $\pi^{+/-}$ and $K^{+/-}$ exhibit 
similar $<\!\!p_{T}\!\!>$ to those in $pp$ collisions, but protons and 
anti-protons have significantly higher $<\!\!p_{T}\!\!>$, indicating 
that nuclear effects are important  even at small $N_{part}$. 
For all measured particles species, $<\!\!p_{T}\!\!>$ increases by
$\approx12-14\% $ from the first to the second centrality bin 
(i.e. $N_{part}$ 14 to 79).  
Above $N_{part}=100$ the $\pi^{+/-}$ and $K^{+/-}$  $<\!\!p_{T}\!\!>$ 
appear to  saturate, whereas the $p$ and $\overline{p}$ $<\!\!p_{T}\!\!>$ 
rises slowly. However, we note that going from peripheral to the 
most central event class the overall increase in $<\!\!p_{T}\!\!>$ is 
$\approx20\pm5\%$ independent of particle species.

Fig.~\ref{f:fig3} shows the yields per participant versus $N_{part}$.
Error bars include statistical and multiplicity 
dependent systematic errors. Systematic uncertainties in $N_{part}$ that 
can move all curves independent of mass are shown with bands around 
the positive hadron  points. Total systematic uncertainties in 
 $N_{part}$ and the yields at each $N_{part}$ are listed in 
Table~\ref{t:table1}.  

For all particle species, the yield per participant increases with $N_{part}$.
As for $<\!\!p_{T}\!\!>$, most of the increase occurs between the two 
most peripheral centrality selections (i.e. $N_{part}$ 14 to 79). However, 
in contrast with the centrality dependence of $<\!\!p_{T}\!\!>$, we see an 
indication that the total increase in the yields per participant differs 
among particle species as we go from peripheral to central events. 
The pion yield per participant rises by $21\%\pm6\%(stat)\pm8\%(syst)$. 
The kaon yields per participant rise faster: 
$94\%\pm11\%(stat)\pm26\%(syst)$ and  
$66\%\pm12\%(stat)\pm20\%(syst)$, for $K^{+}$ and $K^{-}$, respectively. 
Similar trends in the centrality dependence of strangeness production have 
been observed in lower energy  heavy ion collisions at the  
BNL-AGS~\cite{E802_1_2} and at the CERN-SPS~\cite{NA49}.
It is interesting to note that at RHIC 
$p$ and $\overline{p}$ yields per participant behave similarly to the 
$K^{+/-}$ yields and also rise faster than the pions with increasing 
$N_{part}$. The increase is $58\%\pm5\%(stat)\pm16\%(syst)$ and 
$72\%\pm9\%(stat)\pm20\%(syst)$, respectively. The similar centrality 
dependence in $p$ and $\overline{p}$ yields per participant indicates that  
baryon/anti-baryon pair production is the dominant source of protons and 
anti-protons alike.

In heavy ion collisions at AGS energies~\cite{E878_1_2} the 
$\overline{p}$ production is close to threshold, the yields per participant 
are lower than in $pp$ collisions and decrease from peripheral to 
central collisions, probably due to annihilation. At the SPS, 
the $\overline{p}$ yield per participant is larger than the $pp$ value and has 
almost no centrality dependence~\cite{NA49_ap}. At RHIC, the total yield of 
anti-protons at mid-rapidity in central Au-Au collisions is a factor of  
$\simeq 1000$ larger than at the AGS~\cite{E878_1_2} and nearly an order of 
magnitude above that in Pb+Pb collisions at CERN~\cite{NA442}. Most of 
the increase is due to the $\sqrt{s}$ dependence of baryon/anti-baryon pair 
production, however the yield per participant rises noticably 
from peripheral to central collisions. 

In conclusion, an intriguing new behavior in identified hadron 
production at RHIC is reported. In central Au-Au collisions the anti-protons 
yield is comparable to the $\pi^{-}$ at high $p_{T}$ - a behavior never 
observed before in elementary or in heavy-ion collisions. $<\!\!p_{T}\!\!>$ 
rises with centrality similarly for all particle species, while $K^{+/-}$, 
$p$ and $\overline{p}$ yields  per participant increase somewhat faster than 
the $\pi^{+/-}$ yields.

%
%
%
%
%
%
%
%


We thank the staff of the Collider-Accelerator and Physics Departments at
BNL for their vital contributions.  We acknowledge support from the
Department of Energy and NSF (U.S.A.), MEXT and JSPS (Japan), RAS,
RMAE, and RMS (Russia), BMBF, DAAD, and AvH (Germany), VR and KAW
(Sweden), MIST and NSERC (Canada), CNPq and FAPESP (Brazil), IN2P3/CNRS
(France), DAE and DST (India), KRF and CHEP (Korea), the U.S. CRDF for 
the FSU, and the US-Israel BSF.





\begin{table*} 
\caption{Integrated hadron  
( $\pi^{\pm}, K^{\pm}$, $p$ and $\overline{p}$ ) yields 
at mid-rapidity  for five centrality classes (see text) 
identified by the indicated number of participants $N_{part}$. The errors on 
$N_{part}$ are the systematic errors. The errors listed for 
$<\!\!dN/dy_{|y=0}\!\!>$ are statistical. The systematic errors are 
$13\%$, $15\%$ and $14\%$ for $\pi^{\pm}$, $K^{\pm}$, $p$ and $\overline{p}$, 
respectively.\label{t:table1}}
\begin{tabular}[]{ccccccc}
$N_{part}$ &348$\pm$10.0 &271$\pm$8.4 &180$\pm$6.6&79$\pm$4.6 & 14$\pm$3.3\\
\hline
$\pi^{+}$ &276$\pm$3 &216$\pm$2 &141$\pm$1.5 &57$\pm$0.6&9.6$\pm$0.2 \\ 
$\pi^{-}$ &270$\pm$3 &200$\pm$2 &129$\pm$1.4 &53.3$\pm$0.6&8.6$\pm$0.2\\
$K^{+}$ &46.7$\pm$1.5 &35.0$\pm$1.3&22.2$\pm$0.8 &8.3$\pm$0.3&0.97$\pm$0.11\\
$K^{-}$ &40.5$\pm$2.3 &30.4$\pm$1.4&15.5$\pm$ 0.7&6.2$\pm$0.3&0.98$\pm$0.1\\
$p$  &28.7$\pm$0.9 &21.6$\pm$0.6 &13.2$\pm$0.4 &5.0$\pm$0.2&0.73$\pm$0.06\\ 
$\overline{p}$&20.1$\pm$1.0&13.8$\pm$0.6&9.2$\pm$0.4&3.6$\pm$0.1&0.47$\pm$0.05 \\
\end{tabular}
\end{table*}

\end{multicols}    


\begin{references}
\bibitem[*]{Deceased}Deceased     
\bibitem[\dagger]{non-par} 
Not a participating Institution.

\bibitem{PHENIX1}
K.~Adcox~{\it et~al.},
\Journal{\PRL} {86}{3500}{2001};

\bibitem{PHENIX2}
K.~Adcox~{\it et~al.},
\Journal{\PRL} {87}{052301}{2001};
B.B.~Back~{\it et~al.},
\Journal{\PRL} {85}{3100}{2000};
C.~Adler~{\it et~al.},
\Journal{\PRL} {87}{112303}{2001}.

\bibitem{QMproc} e.g. see {\sl Proc. Quark Matter 1984}, ed. K.~Kajantie
   (Springer, Berlin 1985); {\sl Proc. Quark Matter 1987}, eds. H.~Satz,
   H.~J.~Specht, R.~Stock, \Journal{\ZPC}{38}{1-370}{1988}.
\bibitem{Teaney} 
D.~Teaney~{\it et~al.},
\Journal{\PRL} {86}{4783}{2001}; {nucl-th/0110037}.

\bibitem{Kolb} 
P.~Kolb~{\it et~al.},
\Journal{\NPA} {696}{197}{2001}; 
P.~Huovinen~{\it et~al.},
\Journal{\PLB} {503}{58}{2001}

\bibitem{Jurgen} 
J.~Schaffner-Bielich~{\it et~al.},
{hep-ph/0101133; nucl-th/0108048 and references therein}.

\bibitem{Vitev}
I.~Vitev and Miklos Gyulassy, {nucl-th/0104066}.

\bibitem{Ham}
H.~Hamagaki (PHENIX Collaboration), 
\Journal{\NPA} {698}{412c}{2002}.

\bibitem{TEXAS} J.~Velkovska, Proc. CAARI 2000,
AIP {\bf CP576}, 227 (2001).

\bibitem{NIM} J.~T.~Mitchell~{\it et. al.},
\Journal{\NIMA} {482}{498-515}{2002}

\bibitem{PHOBOS2}
B.B.~Back~{\it et~al.},
\Journal{\PRL} {87}{102303}{2001}.

\bibitem{ppg003}
K.~Adcox~{\it et~al.},
\Journal{\PRL} {88}{22301}{2002}

\bibitem{STAR2}
C.~Adler~{\it et~al.},
\Journal{\PRL} {87}{262302}{2001}.

\bibitem{UA1}
C.~Albajar {\it et~al.},
\Journal{\NPB} {335}{261}{1990}.


\bibitem{E866}
L.Ahle~{\it et~al.},
\Journal{\NPA} {610}{139c-152c}{1996}.

\bibitem{NA441}
I.G.~Bearden~{\it et~al.},
\Journal{\PRL} {78}{2080}{1997}.

\bibitem{NA49}
J.~Bachler~{\it et~al.},
\Journal{\NPA} {661}{45c-54c}{1999}.


\bibitem{ISR} B.~Alper {\it et~al.},
\Journal{\NPB} {100}{237-290}{1975}.

\bibitem{FERMILAB}
T.~Alexopoulos~{\it et~al.},
\Journal{\PRD} {48}{984}{1993}.

\bibitem{E802_1_2}
L.~Ahle~{\it et~al.},
\Journal{\PRC} {59}{2173}{1999};
\Journal{\PRC} {60}{044904}{1999}.

\bibitem{E814}
J.~Barrette~{\it et~al.},
\Journal{\PRL} {70}{1763}{1993}


\bibitem{E878_1_2}
D.~Beavis~{\it et~al.},
\Journal{\PRL} {75}{3633}{1995};
M.~Bennett~{\it et~al.},
\Journal{\PRC} {56}{1521}{1997};
L.~Ahle~{\it et~al.},
\Journal{\PRL} {81}{2650}{1998}.

\bibitem{NA49_ap}
G.~I.~Veres,
\Journal{\NPA} {661}{383c-386c}{1999}.

\bibitem{NA442}
I.G.~Bearden~{\it et~al.},
\Journal{\PLB} {388}{431-436}{1996}.


\end{references}
\end{document}